\null

\input epsf
 
 
\magnification 1200

 
\newskip\ttglue

\font\eightrm=cmr8
\font\eighti=cmmi8
\font\eightsy=cmsy8
\font\eightbf=cmbx8
\font\eighttt=cmtt8
\font\eightsl=cmsl8
\font\eightit=cmti8
\font\sixrm=cmr6
\font\sixbf=cmbx6
\font\sixi=cmmi6
\font\sixsy=cmsy6
 
\def \eightpoint{\def\rm{\fam0\eightrm}
\textfont0=\eightrm \scriptfont0=\sixrm \scriptscriptfont0=\fiverm
\textfont1=\eighti \scriptfont1=\sixi   \scriptscriptfont1=\fivei
\textfont2=\eightsy \scriptfont2=\sixsy   \scriptscriptfont2=\fivesy
\textfont3=\tenex \scriptfont3=\tenex   \scriptscriptfont3=\tenex
\textfont\itfam=\eightit  \def\it{\fam\itfam\eightit}%
\textfont\slfam=\eightsl  \def\sl{\fam\slfam\eightsl}%
\textfont\ttfam=\eighttt  \def\tt{\fam\ttfam\eighttt}%
\textfont\bffam=\eightbf  \scriptfont\bffam=\sixbf
 \scriptscriptfont\bffam=\fivebf  \def\bf{\fam\bffam\eightbf}%
\tt \ttglue=.5em plus.25em minus.15em
\setbox\strutbox=\hbox{\vrule height7pt depth2pt width0pt}%
\normalbaselineskip=9pt
\let\sc=\sixrm  \let\big=\eightbig  \normalbaselines\rm
}
\def\a{\alpha}

\def\c{\gamma}
\def\d{\delta}
\def\e{\epsilon}

\def\m{\mu}
\def\n{\nu}

\def\p{\pi}
\def\r{\rho}
\def\s{\sigma}

\def\C{\Gamma}
\def\D{\Delta}
\def\L{\Lambda}

\def\S{\Sigma}

\def\pl{\partial}
\def\ra{\rightarrow}

\def\OO{{\cal O}}
\def\BB{{\cal B}}
\def\bBB{\bar{\cal B}}

\def\GV{{\rm GeV}}

\def\el{{\rm e}}
\def\up{{\rm u}}
\def\do{{\rm d}}
\def\st{{\rm s}}
\def\ch{{\rm c}}

\def\pr{{\rm p}}
\def\ne{{\rm n}}
\def\Nu{{\rm N}}
\def\ha{{\rm h}}
\def\Xx{{\rm X}}

\def\SIMQ{\mathrel{\mathop \sim_{Q^2\ra\infty}}} 
\def\EQz{\mathrel{\mathop =_{z \sim 1}}}
\def\aQ{\bigl(\a_s(Q^2)\bigr)}

 
{\nopagenumbers
 
\line{\hfil CERN-TH/97-206    } 
\line{\hfil SWAT 97/159   }
\line{\hfil hep-ph/9709213 }
\vskip2cm
\centerline{\bf TESTING TARGET INDEPENDENCE OF THE `PROTON SPIN' EFFECT}
\vskip0.8cm
\centerline{\bf  IN SEMI-INCLUSIVE DEEP INELASTIC SCATTERING}
\vskip1.5cm
\centerline{\bf G.M. Shore${}^*$ and G. Veneziano${}^{\dagger}$}
\vskip0.8cm
\centerline{\it ${}^*$ Department of Physics}
\centerline{\it University of Wales Swansea}
\centerline{\it Singleton Park}
\centerline{\it Swansea, SA2 8PP, U.K. }
\vskip0.3cm
\centerline{\it ${}^{\dagger}$ Theory Division}
\centerline{\it CERN}
\centerline{\it CH 1211 Geneva 23, Switzerland}
\vskip1.5cm
\noindent{\bf Abstract}
\vskip0.5cm
\noindent A natural consequence of the composite operator propagator-vertex
description of deep inelastic scattering developed by the authors is 
that the anomalous suppression observed in the flavour singlet contribution
to the first moment of the 
polarised proton structure function $g_1^p$ (the `proton spin' problem)
is not a special property of the proton structure but is a target
independent effect which can be related to an anomalous suppression
in the QCD topological susceptibility. In this paper, it is shown how this
target independent mechanism can be tested in semi-inclusive deep inelastic
scattering in which a pion or D meson carrying a large target energy fraction
$z$ is detected in the target fragmentation region.

\vskip2cm
\line{CERN-TH/97-206     \hfil}
\line{SWAT 97/159    \hfil}
\line{August 1997     \hfil}
 
\vfill\eject }

\pageno = 1
 
\noindent {\bf 1. Introduction}
\vskip0.5cm
The anomalous suppression of the first moment, $\C_1^p$, of the 
polarised proton structure function $g_1^p$ has been the focus of intense
theoretical and experimental activity for nearly a decade.
While it is now generally accepted that the key to understanding this effect
is the existence of the chiral $U(1)$ anomaly in the flavour singlet
pseudovector channel, there are several detailed explanations
reflecting different theoretical approaches to the description of
deep inelastic scattering (DIS) and proton structure.
In this paper, we review one of these -- the composite operator
propagator-vertex (CPV) description of deep inelastic scattering developed
by us in a series of papers[1-4] -- and show how one of its key predictions,
the target independence of the suppression mechanism, can be tested
in future semi-inclusive DIS experiments.

The essence of our approach is the decomposition of structure function
moments into the product of perturbative Wilson coefficients,
non-perturbative but target-independent composite operator propagators,
and vertex functions describing the coupling of these operators to
the target nucleon.

The vertices, which are defined to be `1PI' with respect to a chosen set
of operators, encode all the information on the structure and properties
of the target. They are non-perturbative and not directly calculable,
and play the same role in our formalism as the parton densities in the
conventional QCD parton model description of DIS. However, just as the 
parton densities have a universal character, being equally applicable
to DIS or hadron-hadron scattering, these 1PI vertices also have a 
more universal role, being related in favourable cases to low energy
nucleon couplings such as $g_{\pi NN}$ etc. They provide an
alternative, complementary, description of the nucleon state.

However, the most important feature of the CPV formalism as far as the 
`proton spin' problem is concerned is the separation of the composite
operator propagator from the target-dependent vertex. This allows us
to distinguish between generic non-perturbative properties of QCD
and effects which are characteristic of the particular target. Our 
proposal is that the anomalous suppression in the 
flavour singlet contribution to the first moment of $g_1^p$
is of the first kind, viz.~a generic, {\it target-independent} feature of QCD,
related to the chiral $U(1)$ anomaly but not special to any particular hadron.
In fact[1-3], we are able to relate the relevant propagator to a 
fundamental correlation function in QCD, viz.~the topological susceptibility
$\chi(0)$, and show that the suppression in $\C_1^p$ is due to an
anomalously small value of its first moment $\chi'(0)$.
To confirm this interpretation, we have evaluated $\chi'(0)$ using 
QCD spectral sum rules[4], and have found a suppression in good quantitative
agreement with the current data[5,6] on $g_1^p$.

The natural next step is to see whether this target-independent suppression
mechanism can be tested directly, by studying the structure functions
of other targets besides the proton and neutron.
(Unfortunately, for these two targets, isospin invariance already implies
target independence.)
The obvious choice for an alternative target is the photon, whose structure
function $g_1^\c$ may be measured in two-photon processes at a sufficiently
high-luminosity $\el^+\el^-$ collider. However, this turns out to be
an exceptional case, since there is a direct axial current -- two photon
coupling via the electromagnetic chiral $U(1)$ anomaly.
The first moment sum rule for $g_1^\c(k^2)$, as a function of the target
photon virtuality $k^2$, has been presented in refs.[7,8], together
with estimates of the relevant cross-section asymmetries in polarised
colliders. The dependence of $g_1^\c(k^2)$ on the virtuality displays
many interesting features: $g_1^\c(0)$ is zero by electromagnetic current
conservation[9]; its asymptotic value for $k^2$ greater than the
hadronic scale is essentially given by the electromagnetic anomaly
coefficient, with logarithmic corrections governed by the gluonic anomaly; 
and its detailed dependence on $k^2$ as the various quark thresholds
are crossed depends critically on the realisation of chiral symmetry in QCD.

Direct DIS experiments on other hadronic targets are of course not feasible.
We can nevertheless test our ideas in {\it semi-inclusive} DIS in an 
appropriate kinematic region where the reaction is well described in terms
of deep inelastic photon scattering off a Reggeon (or more complicated
exchanged object) with well-defined hadronic characteristics. 
In particular, using the target-independent suppression
hypothesis, we are able to formulate predictions for ratios of cross section
moments (related to moments of the Reggeon structure functions) which
are significantly and characteristically different from expectations
based, like the Ellis-Jaffe sum rule for $g_1^p$, on the simple
valence quark model or OZI rule. In particular, our target-independent
mechanism should be clearly testable by comparing the ratios of
cross section moments for the semi-inclusive reactions 
$\el \pr \ra \el \pi^- ({\rm D}^-) \Xx$ and 
$\el \ne \ra \el \pi^+ ({\rm D}^0) \Xx$, in which a pion or D meson 
carrying a large target energy fraction $z$ is detected in the target
fragmentation region.

The paper is organised as follows. In section 2, we review the most
important features of the CPV method
and its application to the polarised proton structure function,
explain why it leads to a target-independent suppression,
and compare our prediction for the first moment with the most recent
SMC data. Then, in section 3, we show how by assuming target-independence
and exploiting flavour $SU(3)$ symmetry we can derive predictions for 
ratios of structure function moments for a variety of hadrons, including
some which differ dramatically from results using the OZI rule.
 
Semi-inclusive DIS is introduced in section 4, where we use a
combination of symmetry and dynamical arguments to show that our
predictions for ratios of structure function moments can be realised
as ratios of cross section moments in a certain kinematical region.
In this region, the cross sections may be written in terms of
Reggeon structure functions, where the exchanged Regge trajectory 
has the required $SU(3)$ properties. We then compare 
these results with the more precise description of semi-inclusive
DIS in terms of fracture functions[10], and relate the Reggeon
structure function to the recently introduced extended fracture
functions[11]. We conclude with a summary of our predictions for 
the most interesting ratios of semi-inclusive cross sections.

\vfill\eject
 
\noindent{\bf 2. Target Independence and Composite Operator Propagator-Vertex
Method}
\vskip0.5cm
The starting point is the sum rule for the first moment of the polarised 
structure function $g_1^p$, viz.
$$\eqalignno{
\Gamma^p_1(Q^2) &\equiv
\int_0^1 dx~ g_1^p(x;Q^2) \cr
&= {1\over12} C_1^{\rm NS}\aQ \Bigl( a^3
+ {1\over3} a^8 \Bigr) + {1\over9} C_1^{\rm S}\aQ a^0(Q^2) 
&(2.1) \cr}
$$
Here, $a^3$, $a^8$ and $a^0(Q^2)$ are the form factors in the forward 
proton matrix elements of the renormalised axial current, i.e.
$$\eqalignno{
&\langle p, s|A_{\m }^3|p, s\rangle = s_\m {1\over2} a^3
~~~~~~~~~~
\langle p, s|A_{\m }^8|p, s\rangle = s_\m {1\over{2\sqrt3}} a^8
~~~~~~~~~~
\langle p, s|A_{\m }^0|p, s\rangle = s_\m a^0(Q^2)  \cr
&{} &(2.2) \cr}
$$
where $p_\m$ and $s_\m$ are the momentum and polarisation vector of the
proton. 
The $Q^2$ dependence of the singlet form factor
follows from the renormalisation of the singlet current described below.
The perturbative series $C_1^{\rm NS}\aQ$ and $C_1^{\rm S}\aQ$
are OPE coefficients and are now both known to $O(\a_s^3)$ [12-14].

Because of the chiral $U(1)$ anomaly, the singlet current $A_{\m }^0$
is renormalised and mixes with the topological density. Defining
the bare operators $A_{\m B}^0 = \sum \bar q \c_\m \c_5 q$
and $Q_B = {\a_s\over{8\p}} \e^{\m\n\r\s}{\rm tr} G_{\m\n} G_{\r\s}$, 
we have (for $n_f$ flavours)
$$\eqalignno{
A_{\m }^0 &= Z A_{\m B}^0 \cr
Q &= Q_B - {1\over2n_f}(1-Z) \pl^\m A_{\m B}^0  
&(2.3) \cr}
$$
where $Z$ is a divergent renormalisation constant. The associated anomalous 
dimension $\c$ was first calculated in ref.[15] and is now known to 3 loops[16]. 
Matrix elements of $A_{\m }^0$ therefore have a non-trivial scale dependence 
governed by $\c$. In particular,
$$
{d\over dt}a^0 = \c a^0
\eqno(2.4)
$$
where $t = \ln{Q^2\over\L^2}$. 

The anomalous Ward identities for composite operator propagators
are
$$
\pl^\m \langle0|T ~A_{\m }^0 ~~ \OO|0\rangle
~~-~~2n_f \langle0|T ~Q ~~ \OO|0\rangle
~~=~~\langle \d_A \OO \rangle
\eqno(2.5)
$$
where $\OO$ denotes an arbitrary composite operator and
$\d_A \OO$ is its chiral variation. 
Notice that with these definitions of the renormalised composites,
the combination $\Bigl(\pl^\m A_\m^0 - 2n_f Q\Bigr)$
appearing in the anomalous Ward identities is the same for the bare
or renormalised operators[17]. The possibility of making such a definition
is a consequence of the Adler-Bardeen theorem. 

The sum rule (2.1) is derived using the OPE for two electromagnetic currents.
The dominant contributions arise from the operators of lowest twist
and, within this set, those of spin $n$ contribute to the $n$th moment
of the relevant structure function. Eq.(2.1) is the special case for odd 
parity operators of twist 2 and spin 1, viz.
$$
J^\r(q) J^\s(-q) ~\SIMQ~
2 \e^{\r\s\n\m} {q_\n\over Q^2} 
\biggl[C_1^{\rm NS}\aQ \biggl(A_{\m }^3 + {1\over\sqrt3} A_{\m }^8\biggr)
+ {2\over3} C_1^{\rm S}\aQ A_{\m }^0 \biggr]
\eqno(2.6)
$$

It is at this point that our CPV method
and the conventional parton model analysis of DIS diverge.
In the parton model, the form factors are related to quark and gluon
densities as follows[18]:
$$\eqalignno{
a^3 &= \D\up - \D\do  \cr
a^8 &=  \D\up + \D\do - 2\D\st \cr
a^0(Q^2) &= \D\S - n_f {\a_s(Q^2)\over 2\p} \D g(Q^2) 
&(2.7) \cr}
$$
where $\D\S = \D\up + \D\do + \D\st$
and $\D q = \int_0^1 dx~\bigl(\D q(x) + \D {\bar q}(x)\bigr)$. 
There is a scheme ambiguity in these identifications, which relate four
parton densities to just three measured quantities. The above definitions,
in which $\D\S$ is chosen to be scale invariant to all orders (this is possible 
because of the Adler-Bardeen theorem), are made in the AB factorisation
scheme[18-20].

\vskip0.3cm
\centerline{
{\epsfxsize=4cm\epsfbox{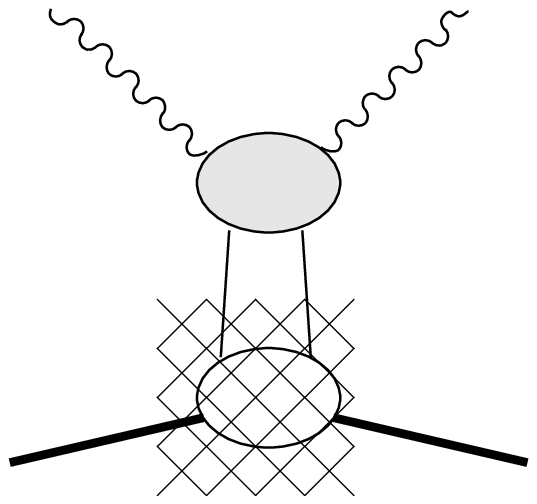}}}
\vskip0.2cm
\noindent{\eightpoint Fig.~1: ~~The description of DIS in the parton model. 
The upper hatched blob denotes the perturbative QCD corrections related to the 
Wilson coefficients in the OPE.}
\vskip0.3cm

This standard approach is illustrated in Fig.~1. The upper hatched blob represents
the perturbative QCD corrections contributing to the coefficient
functions $C_1^{\rm NS}, C_1^{\rm S}$ in the OPE.
The factorisation theorems show that these diagrams,
with two quark (gluon) propagators, give the leading contribution to
the amplitude for large $Q^2$, thus allowing the simple parton interpretation
of $q(x)$ and $g(x,t)$ as the probability distributions for finding a
quark or gluon with momentum fraction $x$ in the target proton.

The Ellis-Jaffe sum rule makes the assumption that $\D\st$ 
and $\D g$ are zero in the proton. This is equivalent to the
OZI (Zweig) rule prediction $a^0 = a^8$. The crux of the `proton spin'
problem is to understand the origin of the OZI breaking revealed by
the measurement of $\C_1^p$, which shows that $a^0$ is strongly
suppressed relative to its OZI value. Our favoured explanation
in the context of the parton model is that the OZI breaking is
due overwhelmingly to the gluon density $\D g$ in eq.(2.7). We expect
the OZI rule to apply to the scale invariant quark densities, so
that (in the AB scheme) $\D\S = a^8$, while the scale dependent
$\D g(Q^2)$ compensates to produce an anomalously suppressed $a^0(Q^2)$.
This would accord with the central conjecture of our rather different approach, 
described below, and has the virtue of providing a scale invariant meaning to 
the OZI rule in the presence of the chiral $U(1)$ anomaly. 
  
In our approach, we again start from the OPE but instead factorise
the resulting matrix elements into the product of composite operator 
propagators and vertex functions.\footnote{$\eightpoint {}^{(1)}$}
{\eightpoint \noindent The presentation here is a slight 
over-simplification. In general, there is a distinction between the cases where 
the OPE operators $\OO_j$ are included in the set $\tilde{\OO}_i$ and where they
are not. See refs.[2,3] for a complete account of the Zumino (partial 
Legendre) transform formalism and its application to the `proton spin' problem.}

\vskip0.3cm
\centerline{
{\epsfxsize=4cm\epsfbox{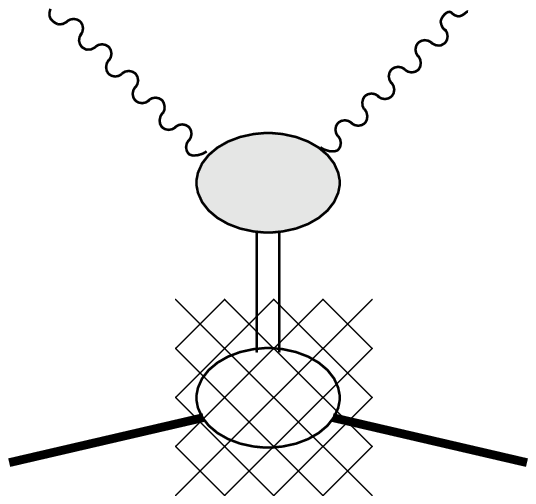}}}
\vskip0.2cm
\noindent{\eightpoint Fig.~2: ~~The description of DIS in the composite operator
propagator-vertex method. The double line denotes the composite operator
propagator and the lower cross-hatched blob the `1PI' vertex function.}
\vskip0.3cm

This is illustrated in Fig.~2.  
To do this, we first select a set of composite operators $\tilde {\OO}_i$
appropriate (see below) to the physical situation and define vertices 
$\C_{\tilde{\OO}_i pp}$ as `1PI' with respect to this set.
Technically, this is achieved by introducing sources for these operators 
in the QCD generating functional, then performing a Legendre transform
to obtain an effective action $\C[\tilde{\OO}_i]$. The 1PI vertices are the 
functional derivatives of $\C[\tilde{\OO}_i]$.
A generic structure function sum rule then takes the form
$$
\int_0^1dx~x^{n-1}~F(x,Q^2) ~=~
\sum_i \sum_j C_j^{(n)}(Q^2) \langle0|T~\OO_j^{(n)} ~\tilde{\OO}_i |0\rangle
\C_{\tilde{\OO}_i pp}
\eqno(2.8)
$$
where $\OO_j^{(n)}$ are the lowest twist, spin $n$, operators in the
appropriate OPE with $C_j^{(n)}$ the corresponding Wilson coefficients.

This decomposition splits the structure function into three pieces -- first,
the Wilson coefficients $C_j^{(n)}(Q^2)$ which control the $Q^2$ dependence and 
can be calculated in perturbative QCD; second, non-perturbative but
{\it target-independent} QCD correlation functions (composite operator
propagators) $\langle0|T~\OO_j^{(n)} ~\tilde{\OO}_i |0\rangle$; and third,
a non-perturbative, target-dependent vertex functions $\C_{\tilde{\OO}_i pp}$
describing the coupling of the target proton to the composite operators of interest.
The vertex functions cannot be calculated directly from first principles.
They encode the information on the nature of the proton state and play an 
analogous role to the parton distributions in the more conventional
parton picture. 

One of the main advantages of our method is that some non-perturbative 
information which is generic to QCD, i.e.~independent of the target, is
factored off into the composite operator propagator. This allows us to 
distinguish between non-perturbative mechanisms which are generic to all
QCD processes and those which are specific to a particular target.
Our contention is that the anomalous suppression in the first moment
of $g_1^p$ is of the first, target-independent, type.

As emphasised in refs.[3,4,21,22], it is important to recognise that this
decomposition of the matrix elements into products of propagators
and proper vertices is {\it exact}, independent of the choice of
the set of operators $\tilde{\OO}_i$. In particular, it is not necessary
for $\tilde{\OO}_i$ to be in any sense a complete set. All that happens if a 
different choice is made is that the vertices $\C_{\tilde{\OO}_i pp}$
themselves change, becoming `1PI' with respect to a different
set of composite fields. Of course, while any set of $\tilde{\OO}_i$ may be
chosen, some will be more convenient than others. Clearly, the set 
of operators should be as small as possible while still capturing the
essential physics (i.e.~they should encompass the relevant degrees of
freedom) and indeed a good choice can result in vertices $\C_{\tilde{\OO}_i pp}$
which are both RG invariant and closely related to low energy physical 
couplings, such as $g_{\p NN}$ or $g_{\p\c\c}$[3,23]. In this case, 
eq.(2.8) provides a rigorous relation between high $Q^2$ DIS and low-energy 
meson-nucleon scattering. 

For the first moment sum rule for $g_1^p$, it is most convenient to use
the chiral anomaly immediately to re-express $a^0(Q^2)$ in terms of the
forward matrix element of the topological density $Q$, i.e.
$$
a^0(Q^2) ~=~{1\over 2M} 2n_f \langle p|Q|p\rangle
\eqno(2.9)
$$
where the matrix element, which scales with the anomalous dimension $\c$,
is evaluated at the scale $Q^2$.\footnote{$\eightpoint {}^{(2)}$}{\eightpoint 
\noindent This quantity may be evaluated directly in lattice QCD. 
See ref.[24] for a brief review of the current status of lattice evaluations.
Note that in order to incorporate fully the effects of the anomaly,
it is necessary[25] to use dynamical fermions.}

Our set of operators $\tilde{\OO}_i$ is then chosen to be the renormalised 
flavour singlet pseudoscalars $Q$ and $\Phi_{5}$ where, up to a vital 
normalisation factor, the corresponding bare operator is 
$\Phi_{5B} = \sum \bar q \c_5 q$.
The normalisation factor[3,4] is chosen such that in the absence of the 
anomaly\footnote{$\eightpoint {}^{(3)}$}{\eightpoint \noindent
To be precise, what is referred to here is the `OZI limit' of QCD,
defined in ref.[25] as the truncation of full QCD in which non-planar
and quark-loop diagrams are retained, but diagrams in which the external
currents are attached to distinct quark loops (so that there are
purely gluonic intermediate states) are omitted. This is a more
accurate approximation to full QCD than either the leading large $1/N_c$ limit,
the quenched approximation (small $n_f$ at fixed $N_c$) or the
leading topological expansion ($N_c\ra\infty$ at fixed $n_f/N_c$.
In the OZI limit, the $U(1)$ anomaly is absent, as is meson-glueball
mixing[26], and there is an extra $U(1)$ Goldstone boson. 
Notice, however, that no approximation is used in deriving eq.(2.10). 
The OZI limit is used here purely as the motivation for choosing a particularly
convenient normalisation for $\Phi_5$.}, $\Phi_{5}$ would have the correct 
normalisation to couple with unit decay constant to the $U(1)$ Goldstone boson 
which would exist in this limit. 
This is important later in justifying the use of the OZI approximation for the
vertex, which is then RG invariant.

We then have
$$
\C_{1~singlet}^p ~=~{1\over9} {1\over2M} 2n_f~
C_1^{\rm S}\aQ \biggl[\langle 0|T~ Q~ Q|0\rangle \C_{Qpp}
~+~\langle 0|T~ Q~ \Phi_{5}|0\rangle \C_{\Phi_5 pp} \biggr]
\eqno(2.10)
$$
where the propagators are at zero momentum and the vertices 
(which in this equation have the external proton wave functions amputated)
are 1PI wrt $Q$ and $\Phi_{5}$ only.

The composite operator propagator in the first term is the zero-momentum
limit of the QCD topological susceptibility $\chi(k^2)$, viz.
$$
\chi(k^2) = \int dx e^{ik.x} i\langle 0|T~ Q(x)~Q(0)|0\rangle
\eqno(2.11)
$$
The anomalous chiral Ward identities show that $\chi(0)$ vanishes 
for QCD with massless quarks, in contrast to pure Yang-Mills theory where
$\chi(0)$ is non-zero. Furthermore, it can be shown[3,4] that the  
propagator $\langle 0|T~ Q~\Phi_{5}|0\rangle$ at zero momentum
is simply the square root of the first moment of the topological
susceptibility. We therefore find:
$$
\C_{1~singlet}^p ~=~ {1\over9} {1\over 2M} 2n_f~
C_1^{\rm S}\aQ ~\sqrt{\chi^{\prime}(0)} ~~\Gamma_{\Phi_{5} pp}
\eqno(2.12)
$$
The quantity $\sqrt{\chi^\prime(0)}$ is
not RG invariant and scales with the anomalous dimension $\gamma$.
On the other hand, the proper vertex has been chosen specifically
so as to be RG invariant. The renormalisation group properties of this
decomposition are crucial to our resolution of the `proton spin' problem.
 
Our proposal (which is fully motivated in refs.[3,23] and supported by
a range of low-energy phenomenology in the $U(1)$ channel, such as
$\eta' \ra \c\c$ decay) is that we should expect the source of OZI violations
to lie in the RG non-invariant, and therefore anomaly-sensitive, 
terms, i.e. in $\chi^{\prime}(0)$.\footnote{$\eightpoint {}^{(4)}$}{\eightpoint
\noindent Notice that we are using RG non-invariance, i.e.~dependence 
on the anomalous dimension $\c$, merely as an indicator of which quantities
are sensitive to the anomaly and therefore likely to show OZI violations.
An alternative suggestion, in which the suppression in $a^0(Q^2)$ is
due directly to non-perturbative effects in $\c$ at low scales, was made in 
ref.~[27]. This would also predict a target-independent suppression.} Since 
the anomalous suppression in $\Gamma_1^p$ is assigned to the composite operator 
propagator rather than the proper vertex, the suppression is a target 
independent property of QCD related to the chiral anomaly, not a special 
property of the proton structure. This immediately raises the question whether 
it is possible to test the mechanism by effectively performing DIS experiments 
on other hadronic targets.

Our quantitative prediction then follows by using the OZI approximation
for the vertex  $\Gamma_{\Phi_{5} pp}$ and a QCD spectral sum 
rule estimate of the first moment of the topological 
susceptibility.\footnote{$\eightpoint {}^{(5)}$}{\noindent \eightpoint
The validity of this calculation has been criticised by Ioffe[28,29] (see
also ref.[30,31]), who asserts that the spectral sum rule technique cannot
be applied to the $U(1)$ channel because of problems with the optimisation
scale and dependence on the strange quark mass. In ref.[32], we 
extend our analysis to include light quark masses and explain in detail why 
these criticisms are not valid.} We find, for $n_f=3$,  
$$
\sqrt{\chi^\prime(0)}\Big|_{Q^2=10 GeV^2}
= 23.2 \pm 2.4~{\rm MeV}
\eqno(2.13)
$$
This is a suppression of approximately a factor $0.6$ relative to the OZI value 
$f_\pi /\sqrt6$. 

Our final result is then
$$
a^0(Q^2=10\GV^2) ~=~ 0.35 \pm 0.05
\eqno(2.14)
$$
from which we deduce
$$
\C_1^p\Big|_{Q^2=10\GV^2} ~=~ 0.143 \pm 0.005
\eqno(2.15)
$$
This is to be compared with the Ellis-Jaffe (OZI) prediction of $a^0 = 0.58
\pm 0.02$ and the SMC experimental data[5]:
$$
\C_1^p\Big|_{Q^2=10\GV^2} ~=~ 0.136 \pm 0.013 \pm 0.009 \pm 0.005
\eqno(2.16)
$$
where the last error is theoretical, related to the $Q^2$ evolution.
This gives
$$
a^0(Q^2=10\GV^2) ~=~ 0.28 \pm 0.16
\eqno(2.17)
$$

There is, however, a remaining uncertainty over the data related to the
small $x$ region. The SMC experiment is limited to measuring the region
$0.003<x<0.7$, and only a small estimated contribution of $0.0042\pm0.0016$
is included in eq.(2.16) for the contribution to $\C_1^p$ from the unmeasured 
range $0<x<0.003$. (The high $x$ extrapolation is uncontroversial.)
Recent fits[19,33] to the same data using a different extrapolation to the 
small $x$ region, incorporating $Q^2$ evolution of the parton distributions, 
suggest a much smaller central value for $a^0$ with larger errors, 
viz. $a^0(Q^2=10\GV^2) = 0.10 {\eightpoint \matrix{+0.17 \cr -0.11 \cr}}$.
Interestingly, these fits also suggest that $\D\S = 0.45 \pm 0.09$,
not too far from the OZI value. 

Very recently, new preliminary proton data has become available from
SMC[6]. This gives
$$
\int_{0.003}^1 dx~g_1^p(x;Q^2=10\GV^2) ~=~ 0.146 \pm 0.006 \pm 0.009 \pm 0.005
\eqno(2.18)
$$
The result for the entire first moment depends on how the extrapolation
to the unmeasured small $x$ region is performed. Using a simple
Regge fit, SMC find $\C_1^p = 0.149 \pm 0.012$ from which $a^0= 0.41\pm 0.11$,
while using a small $x$ fit using perturbative QCD evolution of the parton
distributions they find $\C_1^p = 0.135 \pm 0.016$ which implies 
$a^0 = 0.27 \pm 0.15$ ~(all at $Q^2 = 10\GV^2$). 

Clearly, much more analysis, both theoretical
and experimental, of the small $x$ behaviour of the polarised structure 
functions is required and studying this region will be an important goal of 
future polarised collider experiments at HERA.
Nevertheless, the broad agreement with our prediction (2.14) is very
encouraging and strongly suggests that our interpretation and explicit
calculation of the topological susceptibility are correct.

\vfill\eject
 
\noindent{\bf 3. The $g_1$ Sum Rule for Other Targets}
\vskip0.5cm
In this section, we consider the implications of the target-independent
suppression mechanism for the structure functions of other hadrons,
leaving aside temporarily the question of how this may be realised 
experimentally. 

Our basic prediction\footnote{$\eightpoint {}^{(6)}$}{\eightpoint
\noindent The analogous prediction in the parton model would be
to assume that (in the AB scheme) the RG invariant $\D\S$ would take
its OZI value, ie.~$\D\S \simeq \D\S_{val}$, where $\D\S_{val}$ is
the sum of the valence quark densities. These can be distinguished
from the OZI-violating sea quark densities in semi-inclusive DIS
in the current fragmentation region[34,35].
The gluon contribution $\D G \equiv n_f {\a_s(Q^2)\over2\pi}\D g(Q^2)$
would then be given by 
$$
\D G ~=~ (1-\tilde s(Q^2)) ~\D\S_{val}
$$
where $\tilde s(Q^2)$ is simply (3.2) with the Wilson coefficients
omitted. This has the correct scaling property and ensures a target 
independent suppression in $\C_1$.} 
is that for any hadron, the singlet form factor in 
eq.(2.1) can be substituted by its OZI value multiplied by a universal
(target-independent) suppression factor $s(Q^2)$ determined, up to 
radiative corrections, by the anomalous suppression of the first moment
of the topological susceptibility $\sqrt{\chi'(0)}$. For example,
for a hadron containing only $\up$ and $\do$ quarks, the OZI relation
is simply $a^0 = a^8$, so we would predict:
$$
\C_1^p = {1\over12} C_1^{\rm NS}\aQ \Bigl( a^3 + {1\over3}(1 + 4s) a^8 \Bigr)
\eqno(3.1)
$$
where
$$
s(Q^2) = {C_1^{\rm S}\aQ \over C_1^{\rm NS}\aQ}  ~{a^0(Q^2)\over a^8}
\eqno(3.2)
$$
Since $s$ is target independent, we can use the value measured for the proton 
to deduce $\C_1$ for any other hadron target simply from the flavour 
non-singlet form factors, which obey relations from flavour $SU(3)$ symmetry.
From our spectral sum rule estimate of $\sqrt{\chi'(0)}$, we find 
$s\sim 0.66$ at $Q^2=10{\rm GeV}^2$, while the central value of the SMC 
result (2.17) gives $s\sim 0.55$. (We use the experimental data taken directly 
from SMC, ref.[5] in this section.)

The form factors for a hadron $\BB$ are given by the matrix elements of the
flavour octet axial currents. The $SU(3)$ properties are summarised by
$$
\langle \BB |A_{I I_3 Y}^{\bf (\rho)}|\BB \rangle ~=~
\langle I^\BB I_3^\BB ; I^{\bBB} I_3^{\bBB} | I I_3\rangle ~
\left(\matrix{{\bf \rho}^\BB &{} &{\bf \rho}^{\bBB} &{} &\Big| &{\bf \rho}
&{} \cr I^\BB &Y^\BB &I^{\bBB} &Y^{\bBB} &\Big| &I &Y \cr}\right) ~ 
\langle {\bf \rho}^\BB| A^{\bf (\rho)}|{\bf \rho}^\BB \rangle 
\eqno(3.3)
$$
Here, ${\bf \rho}$ indicates the $SU(3)$ representation while $I, I_3$ and $Y$ 
are the isospin and hypercharge quantum numbers. 
The term $\langle {\bf \rho}^\BB| A^{\bf (\rho)}|{\bf \rho}^\BB\rangle$ 
is a reduced matrix element, while the other factors are $SU(2)$ and $SU(3)$ 
Clebsch-Gordon coefficients[36]. 

If we now take the hadron $\BB$ to be in the {\bf 10} representation,
then since
$$
{\bf 10} \times {\bf \bar{10}} = {\bf 1} + {\bf 8} + {\bf 27} + {\bf 64}
\eqno(3.4)
$$
the matrix element of the (octet) currents contain just one reduced
matrix element. This is in contrast to the case of $\BB$ in the octet 
representation, as for the proton or neutron, which would involve an $F/D$ 
ratio arising from the two reduced matrix elements in the decomposition
${\bf 8} \times {\bf 8} = {\bf 1} + {\bf 8} + {\bf 8} + {\bf 10} 
+ {\bf \bar{10}} + {\bf 27}$. 
This is an important simplification, as it means that the ratio of $\C_1$ for 
decuplet states can be predicted as a simple group-theoretic number, up to 
the dynamical suppression factor $s$.
 
For example, for the $\D^{++}$, the matrix elements of the currents are
$$
\langle \D^{++} | A_\m^3 |\D^{++}\rangle =
\sqrt{{3\over10}} \langle {\bf {10}}|A^{({\bf 8})}|{\bf {10}}\rangle
~~~~~~~~~
\langle \D^{++} | A_\m^8 |\D^{++}\rangle =
\sqrt{{1\over10}} \langle {\bf {10}}|A^{({\bf 8})}|{\bf {10}}\rangle
\eqno(3.5)
$$
evaluating the relevant Clebsch-Gordon coefficients. Similar results hold for 
the $\D^-$. Taking the ratio to eliminate the common reduced matrix element
$\langle {\bf {10}}|A^{({\bf 8})}|{\bf {10}}\rangle$,
we find the following result for the ratio of the first moment of the polarised
structure functions $g_1$ for the $\D^{++}$ and $\D^-$:
$$
{\C_1^{\D^{++}}\over \C_1^{\D^-}} =
{\sqrt{3\over10} + \sqrt{1\over10}\sqrt{1\over3}(1+4s) \over
-\sqrt{3\over10} + \sqrt{1\over10}\sqrt{1\over3}(1+4s)} =
{2s+2\over 2s-1}
\eqno(3.6)
$$
The OZI (c.f.~Ellis-Jaffe) prediction is given by setting $s=1$, i.e.
$\C_1^{\D^{++}}/\C_1^{\D^-} = 4$.\footnote{$\eightpoint {}^{(7)}$}{\eightpoint 
\noindent Alternatively, this result can be simply
obtained in the valence quark model as follows. Using the quark
charges and neglecting radiative corrections, we have
$$
\C_1 = {1\over18} (4\D\up + \D\do + \D\st )
$$
and so 
$$
\C_1^{\D^{++}} = {2\over3}\D\up(\D^{++}) 
~~~~~~~~~
\C_1^{\D^-} = {1\over6} \D\do(\D^-)
$$
With the (isospin) assumption $\D\up(\D^{++}) = \D\do(\D^-)$, we immediately
find $\C_1^{\D^{++}}/\C_1^{\D^-} = 4$.

The corresponding result for the ratio $\C_1^{\S_c^{++}}/\C_1^{\S_c^{0}}$
follows from quark counting in the same way, assuming
$\D\up(\S_c^{++}) = \D\do(\S_c^0)$ and treating the heavy $\ch$ quark
as a spectator.} However, substituting a suppression factor of 
$s \sim 0.66$ gives a much larger ratio $\C_1^{\D^{++}}/\C_1^{\D^-} \sim 10$, 
while the experimental factor $s \sim 0.55$ would give an even larger value,
indicating a near complete suppression of $\C_1^{\D^-}$.

We would therefore expect to find a quite spectacular deviation from 
the quark model expectation for this ratio of structure function 
moments. We can also show (footnote (7)) that the same result is obtained 
for the ratio $\C_1^{\S_c^{++}}/\C_1^{\S_c^{0}}$ for the 
charmed baryons $\S_c^{++} = uuc$ and $\S_c^{0} = ddc$. 
Of course, these examples have been specially selected (because of the $2s-1$ 
factor) to show a particularly striking difference from the simple valence 
quark model predictions. However, as we shall see in section 4, they are also 
the examples which can be dynamically isolated in semi-inclusive DIS.
 
Although these are the most interesting, other ratios of structure function 
moments can be easily calculated by the same method. The most obvious
is the proton-neutron ratio which, as noted above, does not simply give
a group theoretic number but depends also on the $F/D$ ratio.
In this case, the OZI prediction is
$$
{\C_1^p\over\C_1^n} ~=~ {1 - 9F/D \over 4 - 6F/D}
\eqno(3.7)
$$
while including the anomalous suppression factor, we find
$$
{\C_1^p\over\C_1^n} ~=~ {2s -1 -3(2s+1)F/D \over 2s +2 -6sF/D}
\eqno(3.8)
$$
This complements the Bjorken sum rule for the difference $\C_1^p - \C_1^n$, viz.
$$
\C_1^p - \C_1^n ~=~ {1\over6} C_1^{\rm NS}\aQ  ~g_A
\eqno(3.9)
$$
where $g_A = a_p^3$. The neutron structure function is[5]
$$
\C_1^n\big|_{Q^2=10\GV^2} = -0.046 \pm 0.021
\eqno(3.10)
$$
so that the experimental result for the ratio is 
$$
{\C_1^p \over \C_1^n}\bigg|_{Q^2=10\GV^2} = -2.96 \pm 1.39
\eqno(3.11)
$$
This is to be compared with the OZI result  
$\C_1^p/\C_1^n = -7.6 \pm 1.4$, 
where we have used $F/D = 0.575\pm 0.016$ in eq.(3.7),
and with the prediction from our modified formula (3.8) 
which gives central values  
$\C_1^p/\C_1^n = -3.5$ for $s \sim 0.66$ 
and 
$\C_1^p/\C_1^n = -2.9$ for $s \sim 0.55$.
Of course, this only confirms that the suppression in the singlet form factor
$a^0(Q^2)$ is the same for the proton and neutron, as expected by
isospin symmetry and confirmed by the experimental validity of the
Bjorken sum rule.

As a final example, we quote the corresponding results for hadrons 
containing the strange quark. For the octet $\S^+$ and $\S^-$, the
OZI rule gives respectively $a^0 = {1\over2}(3a^3 - a^8)$ and
$a^0 = -{1\over2}(3a^3 + a^8)$, so that our prediction is
$$
\C_1^{\S^+} = {1\over12} C_1^{\rm NS}\aQ 
\Bigl( (1+2s) a^3 + {1\over3} (1-2s) a^8 \Bigr)
\eqno(3.12)
$$
while
$$
\C_1^{\S^-} = {1\over12} C_1^{\rm NS}\aQ 
\Bigl( (1-2s) a^3 + {1\over3} (1-2s) a^8 \Bigr)
\eqno(3.13)
$$
A similar group theoretic calculation then gives
$$
{\C_1^{\S^+} \over \C_1^{\S^-} } ~=~ {2s - 1 - 3(2s+1)F/D \over
2s -1 - 3(2s-1)F/D } 
\eqno(3.14)
$$
with the valence quark model prediction again being recovered by setting $s=1$.

On the other hand, for the $\S^{*+}$ and $\S^{*-}$ in the decuplet, the
$F/D$ ratio is absent and instead we find the simple ratio
$$
{\C_1^{\S^{*+}} \over \C_1^{\S^{*-}} } ~=~ {2s+1 \over 2s-1}
\eqno(3.15)
$$
We shall see to what extent these predictions may be tested in the next section.

\vskip1cm
 
\noindent{\bf 4. Semi-Inclusive Deep Inelastic Scattering}
\vskip0.5cm
Of course, it is not possible to measure structure functions directly
for baryonic targets such as the $\D$ or $\S_c$.
However, it is possible to test the ideas in the previous sections
in semi-inclusive DIS reactions $\el \Nu \ra \el \ha \Xx$, where $\ha$ is a 
detected hadron in the target fragmentation region.

In our case, we are interested in reactions where $\Nu$ is a nucleon target and 
the detected hadron $\ha$ is, for the interesting cases described above, a
pion or D meson. The electron can of course represent any lepton.
There are distinct contributions to this process from the current
and target fragmentation regions, which we require to be clearly
distinguished kinematically. A large rapidity gap is therefore required between
$\ha$ and the inclusive hadrons $\Xx$, with $\ha$ in the target
fragmentation region. $\ha$ is also required to carry a large target energy 
fraction $z$ (defined below).

\vskip0.5cm
\noindent{\bf 4.1 Single Reggeon Exchange Model}
\vskip0.3cm

\centerline{
{\epsfxsize=3.5cm\epsfbox{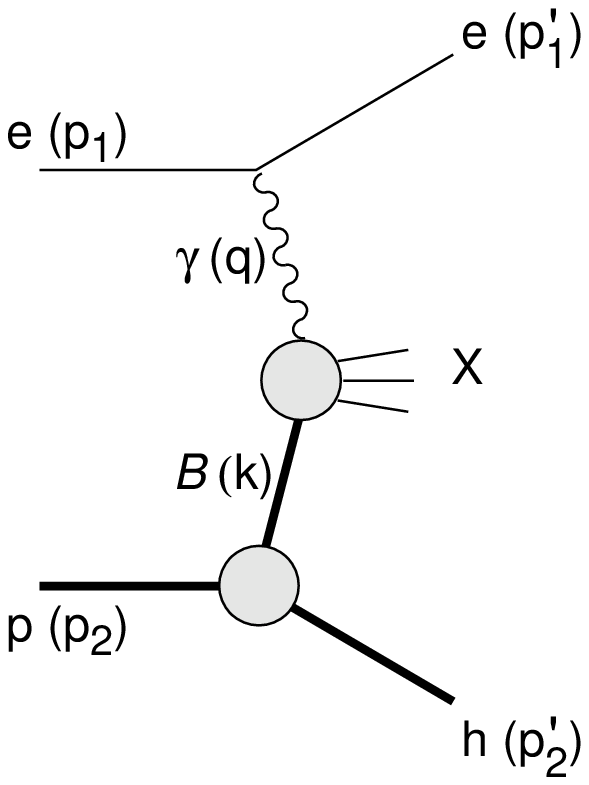}}}
\vskip0.2cm
\noindent{\eightpoint Fig.3~: ~~The semi-inclusive DIS reaction 
$\el \Nu \ra \el \ha \Xx$
in the target fragmentation region with $z \sim 1$ modelled by the
exchange of a Reggeon $\BB$.}
\vskip0.3cm

In this kinematical regime, the process may be modelled as shown in Fig.~3, 
in which the exchanged object is a Reggeon $\BB$ with well-defined $SU(3)$ 
quantum numbers. With a polarised beam and target, this kinematics
allows us to measure the structure function $g_1^{\BB}$ of the 
exchanged Regge trajectory $\BB$.

This is analogous to the measurement of the photon structure function
$g_1^\c$ in polarised $\el^+\el^-$ scattering in the DIS region[7,8]
(Fig.~4) or the pomeron structure function $F_2^{\cal P}$ in diffractive
$\el \pr$ scattering[37] (Fig.~5).
\vskip0.3cm
\centerline{
{\epsfxsize=3cm\epsfbox{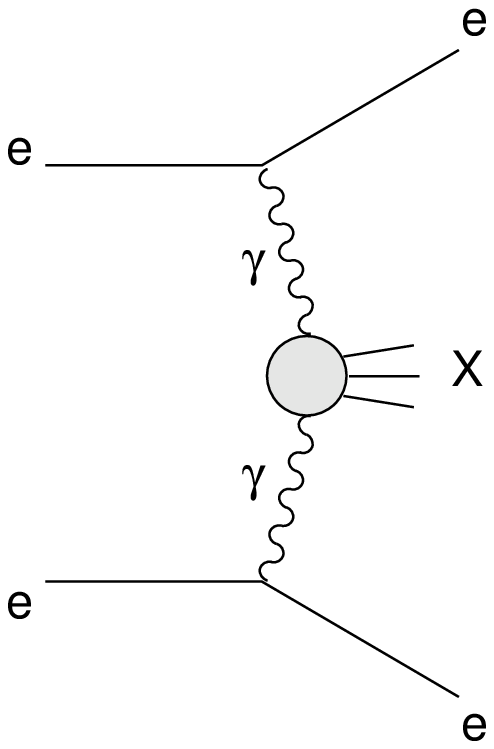}}}
\vskip0.2cm
\noindent{\eightpoint Fig.4~: ~~The deep inelastic two-photon process in polarised 
$\el^+\el^-$ scattering used to measure the structure function $g_1^\c$ 
of the photon.}

\vskip0.3cm
\centerline{
{\epsfxsize=3cm\epsfbox{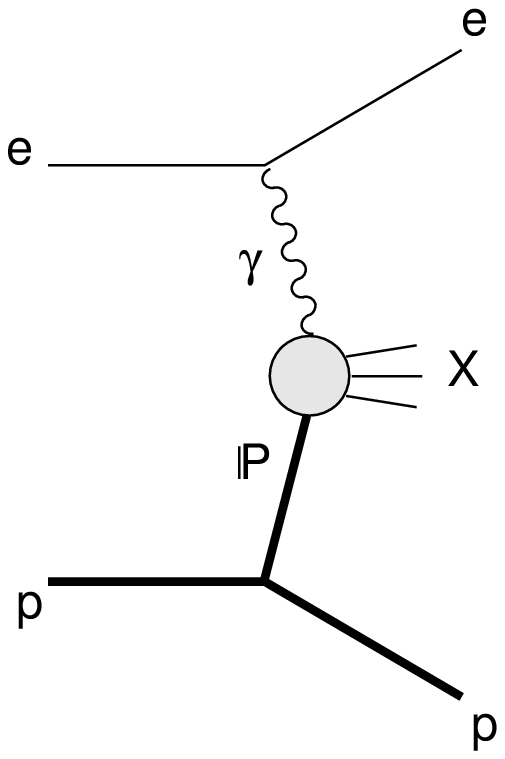}}}
\vskip0.2cm
\noindent{\eightpoint Fig.5~: ~~The diffractive exchange process in $\el\pr$
scattering used to measure the structure function $F_2^{\cal P}$ of 
the pomeron.}
\vskip0.3cm

Since the results of section 3 depend solely on the $SU(3)$ 
properties of the baryon $\BB$, they will still hold here despite
the fact that $\BB$ is interpreted as a Reggeon. 
Indeed, it is not even necessary (see section 4.2) to assume that the
exchanged object is a single Reggeon -- our final predictions for cross section
ratios hold independently of the dynamical nature of the exchanged
object, which could in principle be a multi-Regge exchange or more complicated
structure, provided the $SU(3)$ properties are correct.

If the target is a nucleon $\Nu$ and the detected hadron is an octet meson ($\pi$),
$SU(3)$ symmetry shows that $\BB$ belongs to a representation on 
the rhs of
$$
{\bf 8} \times {\bf 8} = {\bf 1} + {\bf 8} + {\bf 8} + {\bf 10}
+ {\bf \bar{10}} +{\bf 27}
\eqno(4.1)
$$
Since the ${\bf 27}$ requires a 5-quark state, it is a good dynamical
approximation at sufficiently large $z$ that the ${\bf 10}$ dominates the 
${\bf 27}$. However, there is no such argument for ${\bf 8}$ dominance
over the ${\bf 10}$. To isolate a unique representation
for $\BB$, we must therefore choose a combination of $\Nu$ and $\ha$ 
giving $I_3,Y$ quantum numbers for $\BB$ which appear in the 
${\bf 10}$ but not in the ${\bf 8}$. This is 
satisfied by the $\Delta^{++}$ and $\Delta^-$, as in section 3.
The required ratio of first moments $\C_1^{\D^{++}}/
\C_1^{\D^-} = {2s+2\over 2s-1}$, where now $\D^{++}$ 
and $\D^-$ are Reggeons, is therefore obtained by comparing the 
reactions $\el \pr \ra \el \pi^- \Xx$ and $\el \ne \ra
\el \pi^+ \Xx$.

These symmetry considerations are easily pictured by
drawing valence quark diagrams for the $\Nu \ha \BB$ vertex. 
Fig.~6 shows the quark structure of the $\el \pr \ra \el \pi^- \Xx$ 
reaction, while the corresponding 5-quark, ${\bf 27}$ represenation, 
exchange is shown in Fig.~7.

\vskip0.3cm
\centerline{
{\epsfxsize=5cm\epsfbox{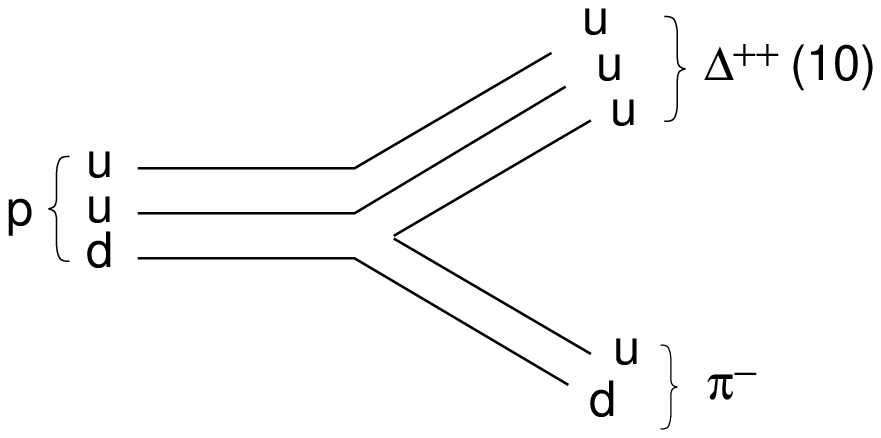}}}
\vskip0.2cm
\noindent{\eightpoint Fig.6~: ~~Quark diagram for the $\Nu\ha\BB$ vertex in the
reaction $\el\pr\ra\el\pi^-\Xx$ with the Reggeon $\BB$ in the {\bf 10}
representation.}

\vskip0.3cm
\centerline{
{\epsfxsize=5cm\epsfbox{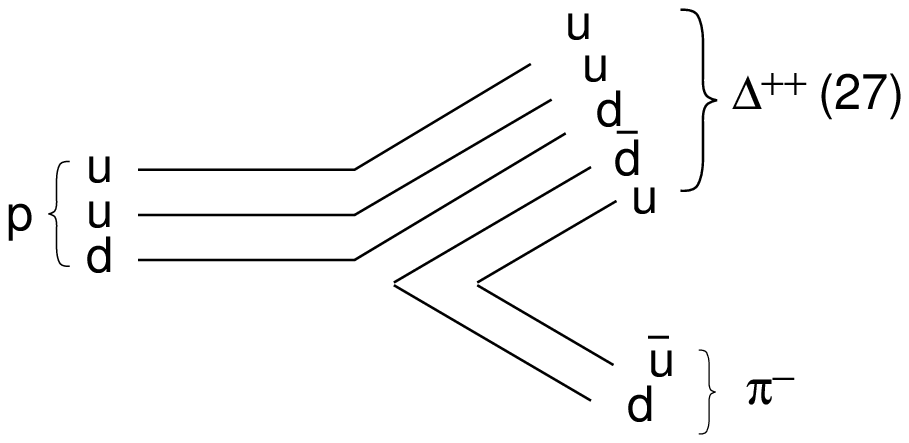}}}
\vskip0.2cm
\noindent{\eightpoint Fig.7~: ~~Quark diagram for the $\Nu\ha\BB$ vertex in the
reaction $\el\pr\ra\el\pi^-\Xx$ with the Reggeon $\BB$ in the {\bf 27}
representation.}
\vskip0.3cm

As in section 3, we find the same ratio holds for the moments 
$\C_1^{\S_c^{++}}/\C_1^{\S_c^{0}}$, which can be realised (Fig.~8)
in reactions in which a D meson is detected by comparing 
$\el \pr \ra \el {\rm D}^- \Xx$ and $\el \ne \ra \el {\rm D}^0 \Xx$.
To justify the assumption made there of treating the $\ch$ quark as
a spectator, we must select events in which there is no charmed jet
in the current fragmentation region.

\vskip0.3cm
\centerline{
{\epsfxsize=5cm\epsfbox{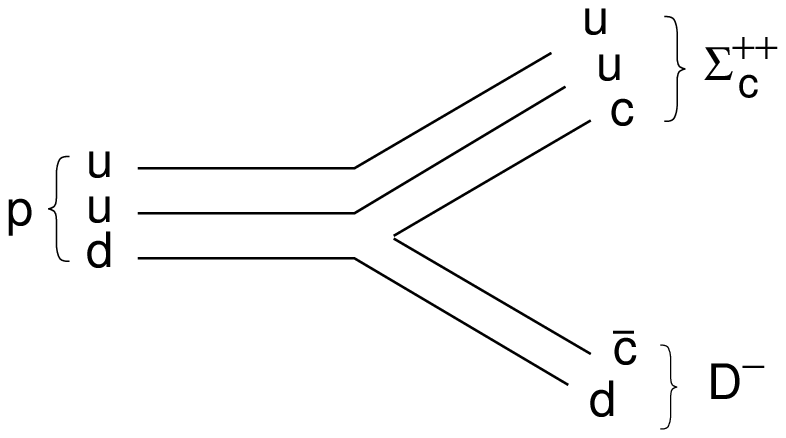}}}
\vskip0.2cm
\noindent{\eightpoint Fig.8~: ~~Quark diagram for the $\Nu\ha\BB$ vertex in the
reaction $\el\pr\ra\el D^-\Xx$ where the Reggeon $\BB$ has quantum numbers
of the $\S_c^{++}$.}
\vskip0.3cm

Trajectories with the quantum numbers of the $\S^+$ and $\S^-$ 
would be exchanged in the reactions $\el \pr \ra \el {\rm K}^0 \Xx$
and $\el \ne \ra \el {\rm K}^+ \Xx$
(substituting an $\st$ quark for the $\ch$ quark in Fig.~8).
However, with these reactions there are two possibilities for the
exchanged trajectory, with either the $\S^+ ~(\S^-)$ in the ${\bf 8}$
or the $\S^{*+} ~(\S^{*-})$ in the ${\bf 10}$ being possible
(in addition to the Zweig suppressed ${\bf 27}$ contribution).
As we saw in section 3, these give quite different ratios of structure
function moments. Unfortunately, it would be difficult to
distinguish these possibilities experimentally. In particular, there
is no dynamical justification for assuming ${\bf 10}$ dominance
over the ${\bf 8}$. This is why we had to choose the $\D^{++} ~(\D^-)$
or $\S_c^{++} ~(\S_c^0)$ trajectories to obtain a clear test of
the predictions of section 3.

The dynamics of these semi-inclusive reactions in the large $z$,
target fragmentation region (Fig.~5) can be deduced by analogy with
the photon structure function or diffractive pomeron exchange processes. 
The first moment of the polarised structure function
for the Reggeon $\BB$ is found from the polarisation asymmetry of the
differential cross section in the target fragmentation region, i.e.
$$
\int_0^{1-z} dx~ x~ {d\D\s^{target}\over dx dy dz dt} 
~=~ {Y_P\over 2} {4\pi \alpha^2\over s} ~\Delta f(z,t)~
\int_0^1 dx_{\BB}~ g_1^{\BB}(x_{\BB},t;Q^2)
\eqno(4.2)
$$
Here, $x = {Q^2\over2p_2.q}$, $x_{\BB} = {Q^2\over2k.q}$, 
$z = {p_2^{\prime}.q \over p_2.q}$ so that $1-z = {x\over x_{\BB}}$, 
$y = {p_2.q\over p_2.p_1}$, $t = -(p_2-p_2^{\prime})^2 \equiv -k^2$
and $Y_P = {1\over y}(2-y)$. This kinematics is described further 
in Appendix A.

If we now take the ratio of eq.(4.2) for the two reactions
$\el \pr \ra \el \pi^- \Xx$ and $\el \ne \ra \el \pi^+ \Xx$ 
(or $\el \pr \ra \el {\rm D}^- \Xx$ and $\el \ne \ra \el {\rm D}^0 \Xx$),
the factorised Reggeon emission factor $\D f(z,t)$ cancels out, leaving 
the ratio of structure function moments $\C_1$ predicted in section 3
to be given simply by the ratio of the cross section moments.
Our final predictions for the cross section ratios are summarised in
section 4.3.

As well as predicting the ratios, which contain the essential physics
we wish to test, we should also consider the absolute size of the
relevant cross section asymmetries. In particular, we must check that the 
cross sections do not fall off too quickly as $z$ approaches 1 for
our predictions to be seen clearly in the data. Returning to eq.(4.2),
and making the ansatz that the Reggeon emission factor $\D f(z,t)$ appropriate
to the polarised amplitude is the same as that for the unpolarised
case, we expect
$$
\D f(z,t) ~\sim~ F(t) (1-z)^{1-2\a_{\BB}(t)}
\eqno(4.3)
$$
where $\a_{\BB}(t)$ is the Regge trajectory for the $\BB$.
For the reactions of interest, the relevant trajectory is the $\D$,
for which $\a_{\D}(t) \simeq 0.0 + 0.9t$, with $t$ in $\GV^2$.
For the relevant experimental condition of $t$ close to zero (corresponding to a 
small scattering angle $\theta_{LAB}$ of $\ha$ relative to the incident nucleon 
in a collider experiment), the cross section moment therefore falls off 
only as $1-z$. 

\vskip0.5cm
\noindent{\bf 4.2 Fracture Functions}
\vskip0.3cm

Within the general framework of the parton model, the appropriate
description of events in the target fragmentation region in semi-inclusive
DIS is with fracture functions, introduced in ref.[10].
In this section, we show briefly how the fracture function description gives
a more rigorous foundation to the results given in the previous section in terms 
of Reggeon structure functions.

\vskip0.3cm
\centerline{
{\epsfxsize=3cm\epsfbox{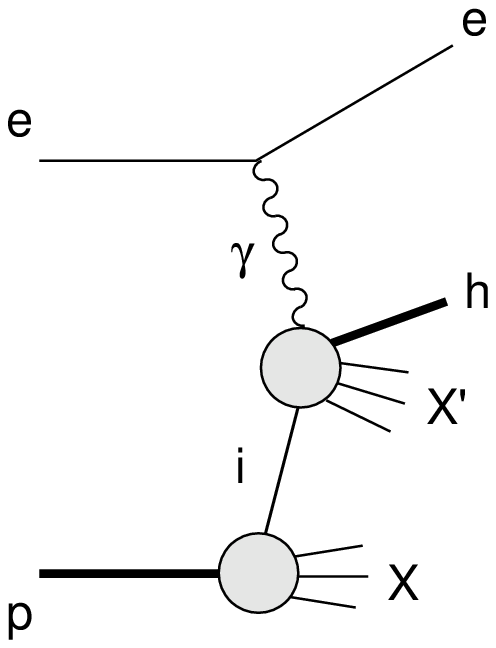}}}
\vskip0.2cm
\noindent{\eightpoint Fig.9~: ~~The contribution to semi-inclusive DIS from the
current fragmentation region.}

\vskip0.3cm
\centerline{
{\epsfxsize=3cm\epsfbox{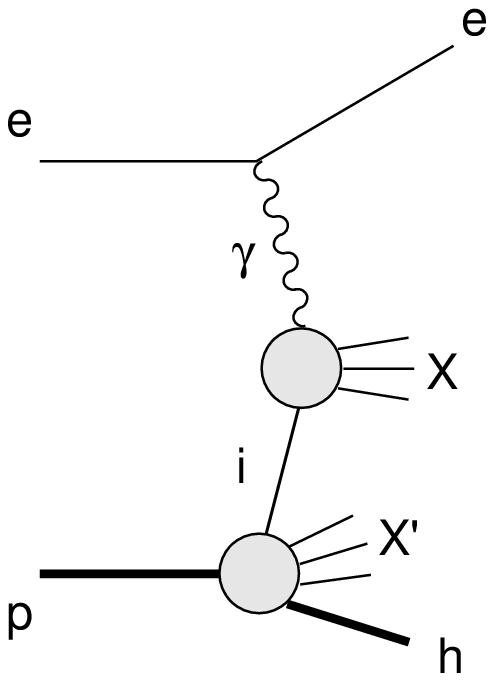}}}
\vskip0.2cm
\noindent{\eightpoint Fig.10~: ~~The contribution to semi-inclusive DIS from the
target fragmentation region.}
\vskip0.3cm

The two distinct contributions to semi-inclusive DIS from the current
and target fragmentation regions are shown in Figs.~9 and 10.
The current fragmentation events are described by parton fragmentation
functions $D_i^h(z;Q^2)$, where $i$ denotes the parton, while the
target fragmentation events are described by fracture functions
$M_i^{hN}(x,z;Q^2)$ representing the joint probability distribution for
producing a parton with momentum fraction $x$ and a detected hadron $\ha$ 
carrying energy fraction $z$ from a nucleon $\Nu$.

The differential cross section for polarised, semi-inclusive DIS has been
given in refs.[38,39], including NLO corrections. (The equivalent
results for the unpolarised case were calculated in ref.[40].)
For our purposes here, we just quote the lowest order result:
$$
x~{d\D\s \over dx dy dz} ~=~ {Y_P\over2} {4\pi\a^2 \over s} ~\sum_i e_i^2
\biggl[{1\over 1-x} \D q_i(x;Q^2) D_i^h\Bigl({z\over 1-x}; Q^2\Bigr)
~+~ \D M_i^{hN}(x,z;Q^2) \biggr]
\eqno(4.4)
$$
where we have expressed the result in terms of the variable 
$z={p_2^{\prime}.q\over p_2.q}$ (see Appendix A).
Here, $\D q_i(x;Q^2)$ and $\D M_i^{hN}(x,z;Q^2)$ are the polarisation
asymmetries of the parton densities and fracture functions respectively.
Restricting to the target fragmentation region, we simply have:
$$
x~{d\D\s^{target} \over dx dy dz} ~=~ 
{Y_P\over2} {4\pi\a^2 \over s} ~\sum_i e_i^2 ~\D M_i^{hN}(x,z;Q^2) 
\eqno(4.5)
$$

In the kinematical region $z \sim 1$ where the dominant process can be 
modelled as Reggeon exchange (Fig.~5), we can compare this
expression to eq.(4.2). In this limit, therefore, we can
relate the Reggeon structure function to the fracture function as follows: 
$$
\sum_i e_i^2 ~\int_0^{1-z} dx ~\D M_i^{hN}(x,z;Q^2)
~~\EQz ~~
\int dt~\D f(z,t) ~\int_0^1 dx_{\BB}~g_1^{\BB}(x_{\BB},t;Q^2)
\eqno(4.6)
$$
This is just the first moment of the more general relation
$$
\sum_i e_i^2 ~\D M_i^{hN}(x,z;Q^2)
~~\EQz ~~
\int dt~{1\over 1-z}~\D f(z,t) ~g_1^{\BB}\Bigl({x\over 1-z},t;Q^2\Bigr)
\eqno(4.7)
$$

This relation expresses the Reggeon structure function $g_1^{\BB}$
in terms of its partonic constituents, as described by the fracture
function $\D M_i^{hN}$. We therefore see that the fracture function
measures the parton distribution of the exchanged object[10].
Indeed, this interpretation is more general than the particular relation 
(4.7), since the fracture function description is not dependent on a 
particular model (such as a single Regge trajectory) for the 
exchanged object. For example, if the process is modelled by multi-Regge
exchange, the rhs of eq.(4.7) would comprise a sum over
the Reggeons.

We can take this identification a stage further by considering the 
extended fracture functions $M_i^{hN}(x,z,t;Q^2)$ introduced
recently in ref.[11]. These are defined such that
$$
M_i^{hN}(x,z;Q^2) ~=~ \int_0^{O(Q^2)} dt~M_i^{hN}(x,z,t;Q^2)
\eqno(4.8)
$$
where $t= -(p_2 - p_2^{\prime})^2$. Just as in the integrals 
of the Reggeon emission factors, the upper limit of the $t$ integration
is not precisely specified, with the physical results for large $Q^2$ 
being independent of the exact choice to the required order.
These extended fracture functions have a number of important features[11],
notably a much simpler, homogeneous, RG evolution equation. They also
have an interesting interpretation in terms of spacelike cut vertices,
whose RG properties are known to be determined by the anomalous
dimensions of appropriate local operators.

These extended fracture functions allow us to remove the $t$ integration
in eq.(4.7), so we can finally write, to leading order,
$$
\sum_i e_i^2 \D M_i^{hN}(x,z,t;Q^2)
~~\EQz ~~ 
F(t) (1-z)^{-2\a_{\BB}(t)} ~g_1^{\BB}\Bigl({x\over 1-z},t;Q^2\Bigr)
\eqno(4.9)
$$
where we have substituted eq.(4.3) for $\D f(z,t)$. The NLO corrections
to the rhs of eq.(4.9) can be read off from refs.[38,39].
In fact, this should be taken as the definition of the `Reggeon structure
function' in the single Regge exchange approximation.

Since $\BB$ is a Reggeon, we cannot express the structure function
$g_1^{\BB}$ in terms of Wilson coefficients and operator matrix elements
as for single particle structure functions such as $g_1^p$.
Nevertheless, the moments of $g_1^{\BB}$ should inherit a RG scaling
dependendence on the anomalous dimension of the appropriate OPE operator,
e.g.~$A_\m^0$ for the flavour singlet first moment. For consistency,
therefore, we would require the extended fracture function 
$\D M_i^{hN}(x,z,t;Q^2)$ to satisfy a homogeneous RG evolution equation.
This is borne out by the results of ref.[11], where it is shown that
$$
{\pl\over\pl\ln Q^2}~\D M_i^{hN}(x,z,t;Q^2)   ~~=~~ {\a_s(Q^2)\over2\pi}~
\int_x^{1-z} {dw\over w} ~\D P_{ij}\Bigl({x\over w}, \a_s(Q^2)\Bigr)~
\D M_j^{hN}(w,z,t;Q^2)
\eqno(4.10)
$$
where $\D P_{ij}$ is the usual DGLAP evolution kernel.

A satisfying picture therefore emerges, in which the results of section 4.1
are confirmed and placed in a broader theoretical framework for the
description of semi-inclusive DIS.

\vskip0.5cm
\noindent{\bf 4.3 Predictions}
\vskip0.3cm

Of course, the ratio (3.6) is only obtained in the limit as $z$ approaches 1, where
the reaction $\el \Nu \ra \el \ha \Xx$ is dominated by the process
in which most of the target energy is carried through into the 
final state $\ha$ by a single quark (see Figs.~6-8).

At the opposite extreme, for
$z$ approaching 0, the detected hadron carries only a small fraction of the 
target nucleon energy and has no special status compared to the other
inclusive hadrons $\Xx$. In this limit, therefore, the ratio of 
cross section moments (4.2) for 
$\el \pr \ra \el \pi^- \Xx$ and $\el \ne \ra \el \pi^+ \Xx$ 
is simply the ratio of the structure function moments for the proton and
neutron, i.e. $\C_1^p/\C_1^n$ as given in section 3.
The same result would hold for the ratio of
$\el \pr \ra \el {\rm D}^- \Xx$ and $\el \ne \ra \el {\rm D}^0 \Xx$.

We therefore predict the following results for the ratios of the differential
cross section moments $\int_0^{1-z} dx~ x~ {d\D\s^{target}\over dx dy dz dt}$~
:
$$\eqalignno{
{\el \ne \ra \el \pi^+ ({\rm D}^0) \Xx \over 
\el \pr \ra \el \pi^- ({\rm D}^-) \Xx} ~~&\sim~~
{2s-1\over2s+2} ~~~~~~~~~~~~~~~~~~~~~~~~~~~(z\ra 1) \cr
&\sim~~ {2s+2-6sF/D \over 2s-1-3(2s+1)F/D} ~~~~~(z\ra 0) 
&(4.11) \cr }
$$

Between these limits, we can only interpolate. We therefore expect
a plot of the ratios of  
$\int_0^{1-z} dx~ x~ {d\D\s^{target}\over dx dy dz dt}$
over the range $0<z<1$ for $\el \ne \ra \el \pi^+ ({\rm D}^0) \Xx$ 
and $\el \pr \ra \el \pi^- ({\rm D}^-) \Xx$ to look like the
sketch in Fig.~11, where the solid line shows the ratios predicted
by (4.11) with $s\sim 0.66$, contrasted with
the ratios predicted by the OZI rule, i.e.~$s=1$ (dotted line).

\vskip0.3cm
\centerline{
{\epsfxsize=6cm\epsfbox{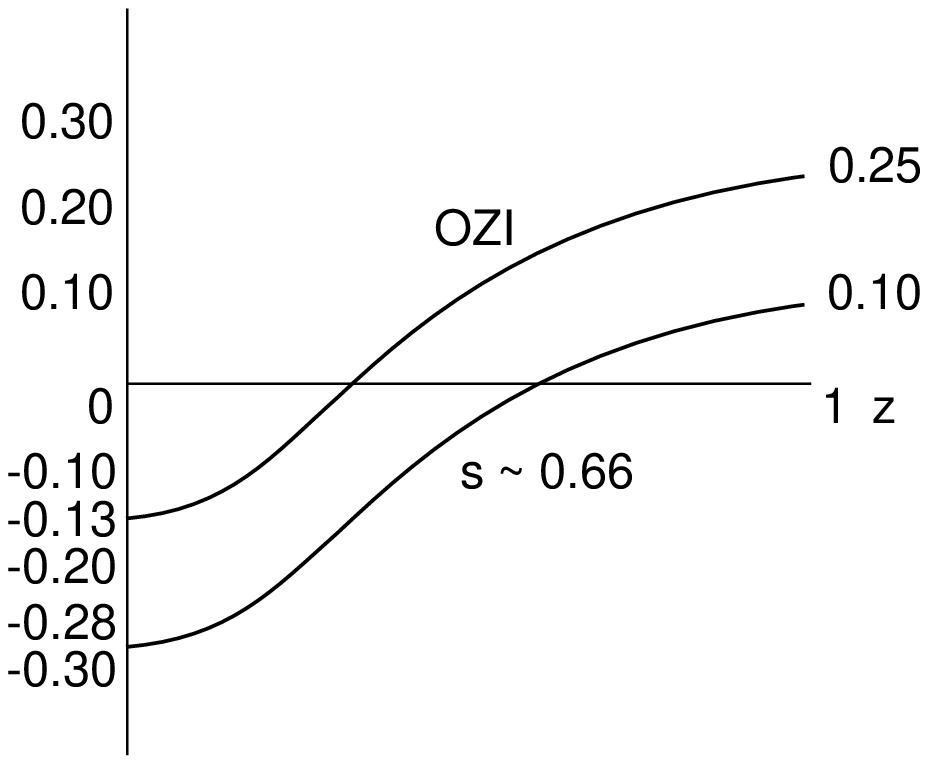}}}
\vskip0.2cm
\noindent{\eightpoint Fig.11~: ~~Sketch of the cross section moment ratios for
$\el \ne \ra \el \pi^+ ({\rm D}^0) \Xx$ and 
$\el \pr \ra \el \pi^- ({\rm D}^-) \Xx$, interpolating between the limits
$z\ra 0$ and $z\ra 1$. The dotted line shows the OZI prediction and the
solid line our prediction based on the target-independent suppression
mechanism (with $s \sim 0.66$).}
\vskip0.3cm

The difference between the OZI (or valence quark model) expectations 
and these predictions based on our target-independent interpretation
of the `proton spin' data is therefore quite dramatic, and should
give a clear experimental signal.

Since our proposed experiment requires particle identification in the
target fragmentation region, it is difficult to do at a polarised 
fixed-target experiment such as COMPASS[41] at CERN, which is better suited
to studying semi-inclusive processes in the current fragmentation region.
A better option is a polarised $\el \pr$ collider, such as HERA[42]. 
Testing our predictions requires comparision of proton and neutron data,
which can be extracted from experiments with polarised deuterons
replacing the protons in the collider.

\vskip1cm
 
\noindent{\bf Acknowledgements}
\vskip0.5cm
The work of one of us (GMS) is partially supported by the EC TMR Network Grant
FMRX-CT96-0008. We would like to thank R.~Ball, D.~de Florian, S.~Forte, 
M.~Grazzini, E-M.~Kabuss, G.~Mallot, S.~Narison and L.~Trentadue
for helpful discussions.

\vfill\eject

\noindent{\bf Appendix A}
\vskip0.5cm

A number of different definitions of the variable `$z$' in semi-inclusive
reactions are used in the literature. Here, we describe the relation between
our notation and that used elsewhere, and present a number of useful
kinematical results.

In refs.[38-40], the kinematics is described in the CM frame of the
virtual photon and target nucleon. Let $E_h$ and $E_N$ be the
energy of the detected hadron and nucleon in this frame and
$\theta$ be the angle between the corresponding momenta.
With the definition[40] ${v} ={1\over2}(1-\cos\theta)$, we see that
the target fragmentation region is characterised by ${v} \sim 1$,
while the current fragmentation region is ${v} \sim 0$.
The hadron energy fraction variable used by ref.[40] is then
$$
z_{(G)} = {E_h\over E_N} {1\over 1-x}
\eqno(A.1)
$$
In contrast, the corresponding variable used in refs.[34,35] is
$$
z_h = {p_2.p_2^{\prime}\over p_2.q}
\eqno(A.2)
$$
The relation is
$$
z_h = z_{(G)} (1-{v})
\eqno(A.3)
$$
Notice[40] that these two variables are approximately equal in the
current fragmentation region but differ substantially in the target 
fragmentation region, where $z_h$ is small.

In terms of the variable $t = -(p_2-p_2^{\prime})^2$, which in the
model of Fig.~5 is the invariant spacelike momentum $-k^2$ of the exchanged
Reggeon, we have
$$
z_h = {xt \over Q^2}
\eqno(A.4)
$$
so that at fixed $x, Q^2$ in the target fragmentation region, $z_h$ is 
simply a measure of $t$. The angle $\theta$ (assuming $t$ is small
compared to $Q^2$) is given by
$$
\theta^2 \simeq {4z\over x(1-x)} {t\over Q^2}
\eqno(A.5)
$$

Our preferred variable $z = {p_2^{\prime}.q\over p_2.q}$ can be expressed
in this frame as
$$
z ~=~ {E_h\over E_N}-{xt\over Q^2} ~=~ (1-x)z_{(G)} + O\Bigl({t\over Q^2}\Bigr)
\eqno(A.6)
$$
so for relatively small $t$, $z$ is simply given by the ratio of the detected 
hadron energy to the target nucleon energy in the photon-nucleon CM frame.
The required kinematical region, where the semi-inclusive reaction is well
approximated by the Reggeon exchange diagram and our prediction for the
cross section moment ratios holds, is therefore ${v} \sim 1$
and $z$ approaching 1.

\vfill\eject

\noindent{\bf References}
\vskip0.5cm
 
\settabs\+\ [&123456] &Author  \cr

\+\ [&1] &G.~Veneziano, Mod.~Phys.~Lett. A4 (1989) 1605 \cr 
\+\ [&2] &G.M.~Shore and G.~Veneziano, Phys.~Lett. B244 (1990) 75 \cr
\+\ [&3] &G.M.~Shore and G.~Veneziano, Nucl.~Phys. B381 (1992) 23 \cr
\+\ [&4] &S.~Narison, G.M.~Shore and G.~Veneziano, Nucl.~Phys. B433 (1995) 209\cr
\+\ [&5] &SMC Collaboration, hep-ex/9702005    \cr
\+\ &{}  &E143 Collaboration, Phys.~Lett. B364 (1995) 61  \cr
\+\ [&6] &SMC Collaboration (E-M.~Kabuss), {\it to be published in} Proceedings, 
QCD97 Montpellier \cr
\+\ &{}  &SMC Collaboration (A.~Magnon), {\it to be published in} Proceedings, 
XVIII International \cr 
\+\ &{} &~~~~~~~~~~Symposium on Lepton-Photon Interactions, Hamburg, July 1997\cr
\+\ [&7] &S.~Narison, G.M.~Shore and G.~Veneziano, Nucl.~Phys. B391 (1993) 69\cr
\+\ [&8] &G.M.~Shore and G.~Veneziano, Mod.~Phys.~Lett. A8 (1993) 373 \cr
\+\ [&9] &S.D.~Bass, Int.~J.~Mod.~Phys. A7 (1992) 6039 \cr
\+ [1&0] &L.~Trentadue and G.~Veneziano, Phys.~Lett. B323 (1994) 201  \cr
\+ [1&1] &M.~Grazzini, L.~Trentadue and G.~Veneziano, {\it in preparation} \cr
\+ [1&2] &S.G.~Gorishny and S.A.~Larin, Phys.~Lett. B172 (1986) 109 \cr
\+ [1&3] &S.A.~Larin and J.A.M.~Vermaseran, Phys.~Lett. B259 (1991) 345 \cr
\+ [1&4] &S.A.~Larin, T.~van Ritbergen and J.A.M.~Vermaseran, Phys.~Lett. 
B404 (1997) 153 \cr
\+ [1&5] & J.~Kodaira, Nucl.~Phys. B165 (1980) 129 \cr
\+ [1&6] &S.A.~Larin, Phys.~Lett. B334 (1990) 1 \cr 
\+ [1&7] &D. Espriu and R. Tarrach, Z. Phys. C16 (1982) 77 \cr
\+ [1&8] &G. Altarelli and G.G. Ross, Phys. Lett. B212 (1988) 391 \cr
\+ [1&9] &R.D.~Ball, S.~Forte and G.~Ridolfi, Phys.~Lett. B378 (1996) 255 \cr
\+ [2&0] &R.D.~Ball, {\it in} Proceedings, Ettore Majorana International School
of Nucleon Structure, \cr
\+\ &{} &~~~~~~~~~~Erice, 1995; hep-ph/9511330 \cr
\+ [2&1] &G.M.~Shore, Nucl.~Phys. B (Proc.~Suppl.) 39B,C (1995) 101 \cr
\+ [2&2] &G.M.~Shore, Nucl.~Phys. B (Proc.~Suppl.) 54A (1997) 122 \cr
\+ [2&3] &G.M.~Shore and G.~Veneziano, Nucl.~Phys. B381 (1992) 3 \cr
\+ [2&4] &G.M.~Shore, {\it to be published in} Proceedings, QCD97 Montpellier \cr
\+ [2&5] &G.~Veneziano, {\it in} `From Symmetries to Strings: Forty Years of
Rochester Conferences', \cr
\+\ &{} &~~~~~~~~~~ed. A.~Das, World Scientific, 1990 \cr
\+ [2&6] &O.~Alvarez and G.M.~Shore, Nucl.~Phys. B213 (1983) 327 \cr
\+ [2&7] &R.D.~Ball, Phys.~Lett. B266 (1991) 473 \cr
\+ [2&8] &B.L.~Ioffe and A.Yu.~Khodzhamiryan, Yad.~Fiz. 55 (1992) 3045 \cr
\+ [2&9] &B.L.~Ioffe, {\it in} Proceedings, Ettore Majorana International
School of Nucleon Structure, \cr
\+\ &{} &~~~~~~~~~~Erice, 1995; hep-ph/9511401 \cr
\+ [3&0] &B.V.~Geshkenbein and B.L.~Ioffe, Nucl.~Phys. B166 (1980) 340 \cr
\+ [3&1] &V.A.~Novikov, M.A.~Shifman, A.I.~Vainshtein and V.I.~Zakharov, \cr
\+\ &{} &~~~~~~~~~~Nucl.~Phys. B191 (1981) 301 \cr
\+ [3&2] &S.~Narison, G.M.~Shore and G.~Veneziano, {\it in preparation} \cr
\+ [3&3] &G.~Altarelli, R.D.~Ball, S.~Forte and G.~Ridolfi, hep-ph/9701289 \cr
\+ [3&4] &L.L.~Frankfurt {\it et al.}, Phys.~Lett.~ B230 (1989) 141 \cr
\+ [3&5] &SMC collaboration, Phys.~Lett. B369 (1996) 93  \cr
\+ [3&6] &J.J.~De Swart, Rev.~Mod.~Phys. 35 (1963) 916  \cr
\+ [3&7] &T.~Gehrmann and W.J.~Stirling, Z.~Phys. C70 (1996) 89  \cr
\+ [3&8] &D.~de Florian {\it et al.}, Nucl.~Phys. B470 (1996) 195 \cr
\+ [3&9] &D.~de Florian, C.A.~Garc\'ia Canal and R.~Sassot, Phys.~Lett.
B389 (1996) 358 \cr
\+ [4&0] &D.~Graudenz, Nucl.~Phys. B432 (1994) 351 \cr
\+ [4&1] &COMPASS collaboration, CERN-SPSLC-96-14 \cr    
\+ [4&2] &G.~Ingelman, A.~De Roeck and R.~Klanner (eds.), `Future Physics at HERA',\cr
\+\ &{} &~~~~~~~~~~Proc., 1996 HERA Workshop \cr

\bye